\documentclass[final]{svjour2}
\usepackage{graphicx}
\usepackage{rotating}
\usepackage{amssymb}
\usepackage{mathptmx}
\usepackage[numbers]{natbib}
\makeatletter
\journalname{Journal of Low Temperature Physics}

\bibpunct{}{}{,}{s}{}{,}

\begin{document}

\newcommand{\hdblarrow}{H\makebox[0.9ex][l]{$\downdownarrows$}-}
\title{The NIKA2 instrument, a dual-band kilopixel KID array for millimetric astronomy }

\author{
M. Calvo$^1$,
A.~Beno\^it$^1$,
A.~Catalano$^{1,2}$,
J.~Goupy$^1$,
A.~Monfardini$^{1,2}$,
N.~Ponthieu$^{1,3}$,
E.~Barria$^1$, 
G.~Bres$^1$, 
M.~Grollier$^1$, 
G.~Garde$^1$, 
J.-P.~Leggeri$^1$, 
G.~Pont$^1$, 
S.~Triqueneaux$^1$,
R.~Adam$^2$,
O.~Bourrion$^2$,
J.-F.~Mac\'ias-P\'erez$^2$,
M.~Rebolo$^2$,
A.~Ritacco$^2$,
J.-P.~Scordilis$^2$,
D.~Tourres$^2$,
C.~Vescovi$^2$,
F.-X.~D\'esert$^3$,
A.~Adane$^4$,
G.~Coiffard$^4$,
S.~Leclercq$^4$,
S.~Doyle$^5$,
P.~Mauskopf$^{5,6}$,
C.~Tucker$^5$,
P.~Ade$^5$
P.~Andr\'e$^7$,
A.~Beelen$^8$,
B.~Belier$^{9}$,
A.~Bideaud$^5$,
N.~Billot$^{10}$,
B.~Comis$^2$,
A.~D'Addabbo$^{11}$,
C.~Kramer$^{10}$,
J.~Martino$^8$,
F.~Mayet$^2$,
F.~Pajot$^8$,
E.~Pascale$^5$,
L.~Perotto$^2$,
V.~Rev\'eret$^7$,
L.~Rodriguez$^7$,
G.~Savini$^{12}$,
K.~Schuster$^4$,
A.~Sievers$^{10}$,
R.~Zylka$^4$
}

\institute{
$^1$
Institut N\'eel, CNRS and Universit\'e de Grenoble, France
\\
$^2$
Laboratoire de Physique Subatomique et de Cosmologie,
Universit\'e Grenoble-Alpes, 
  CNRS/IN2P3, 
  53, rue des Martyrs, Grenoble, France
\\
$^3$
Institut de Plan\'etologie et d'Astrophysique de Grenoble (IPAG), CNRS and Universit\'e de
Grenoble, France
\\
$^4$
Institut de RadioAstronomie Millim\'etrique (IRAM), Grenoble, France
\\
$^5$
Astronomy Instrumentation Group, University of Cardiff, UK
\\
$^6$
Arizona State University, Tempe, AZ, USA
\\
$^7$
Laboratoire AIM, CEA/IRFU, CNRS/INSU, Université Paris Diderot, CEA-Saclay, 91191 Gif-Sur-Yvette, France 
\\
$^8$
Institut d'Astrophysique Spatiale (IAS), CNRS and Universit\'e Paris Sud, Orsay, France
\\
$^9$
Institut d'Electronique Fondamentale (IEF), Universit\'e Paris Sud, Orsay, France
\\
$^{10}$ 
Institut de RadioAstronomie Millimetrique (IRAM), Granada, Spain
\\
$^{11}$
Istituto Nazionale di Fisica Nucleare, Laboratori Nazionali del Gran Sasso, Assergi (AQ), Italy
\\
$^{12}$
University College London, Department 
of Physics and Astronomy, Gower Street, London WC1E 6BT, UK\\
\email{martino.calvo@grenoble.cnrs.fr}
}



\maketitle

\begin{abstract}

NIKA2 (New IRAM KID Array 2) is a camera dedicated to millimeter wave astronomy based upon kilopixel arrays of Kinetic Inductance Detectors\cite{Day03} (KID). The pathfinder instrument, NIKA\cite{Monfardini11}, has already shown state-of-the-art detector performance. NIKA2 builds upon this experience but goes one step further, increasing the total pixel count by a factor $\sim$10 while maintaining the same per pixel performance. For the next decade, this camera will be the resident photometric instrument of the Institut de Radio Astronomie Millimetrique (IRAM) 30m telescope in Sierra Nevada (Spain). 
In this paper we give an overview of the main components of NIKA2, and describe the achieved detector performance. The camera has been permanently installed at the IRAM 30m telescope in October 2015. It will be made accessible to the scientific community at the end of 2016, after a one-year commissioning period. When this happens, NIKA2 will become a fundamental tool for astronomers worldwide.

\if
The NIKA (New IRAM KID Array) camera has been a pathfinder instrument for the Kinetic Inductance Detectors community, being the first camera to perform on-sky observation using a multiplexed readout of KID arrays. Building on this experience, we are now developing the next-generation camera, NIKA2, that will be installed at the IRAM 30m telescope in Sierra Nevada (Spain).

NIKA2 will be equipped with a total of 3 arrays: one for the 2mm band, and 2 for the 1.25mm band, one for each polarization, in order to provide polarization-sensitive observations at this wavelength. The large available correct Field of View (6.5arcmin) will be fully sampled using LEKID based on the Hilbert geometry. Each array will therefore have more than 1000 pixels. The Hilbert LEKID have already been tested for astronomical observation in NIKA, and have shown sate-of-the-art performance approaching the photon-noise limit.

The cryostat used to cool down the arrays is based on a dilution refrigerator coupled to two Pulse Tube coolers. The system can thus work continuously, without the need for recycling, and can be fully operated remotely. The readout of the pixels signal will be done using the Frequency Domain Multiplexed strategy typical of KID detectors. A dedicated board called NIKEL has been developed to this scope.

When fully operational, this ground-breaking instrument will represent a unique tool for the astronomers, with many potential fields of applications. These include for example the detailed mapping of the SZ effect in cluster of galaxies or the study of star-forming regions.

We report the current status of the NIKA2 development, and outline the future steps that we will take before installing the instrument in its final configuration at the IRAM telescope. The installation is planned for September 2015.
\fi


\end{abstract}

\section{Introduction}

The IRAM 30m antenna is on of the largest and most sensitive single dish telescope for millimeter wavelengths currently operating worldwide. Its cabin traditionally hosts both heterodyne receivers, ideally suited for high resolution spectroscopy, and broad-band continuum photometric instruments, which are used to detect extremely faint sources.
The 30m primary dish results in a diffraction limited resolution of 17 and 10arcsec for the 150 and 260GHz bands respectively. Thus, in order to fully sample the 6.5arcmin correct FoV, a photometric camera for this telescope needs arrays containing $\sim10^3$ pixels, having an intrinsic noise comparable to, or below, the photon noise expected at the telescope site. Only recently the developments in the field of cryogenic detectors have made arrays of this type possible.

Coupling high sensitivities to their intrinsic suitability for frequency domain multiplexed readout, KID are the ideal candidates for this kind of application. This is why they have been the technology of choice when the project to build the pathfinder instrument, NIKA, started in 2008.
NIKA has been an extremely successful instrument, being the first camera to conduct multiplexed on-sky observations using KID and leading to outstanding scientific results\cite{Adam}\cite{Cata}. In its final configuration, NIKA counted a total of 356 pixels split over the two bands. Its success has been a key step for the approval of the NIKA2 project as the camera of choice for the IRAM telescope.
\section{The NIKA2 instrument}
While inheriting a lot of know-how from the experience gained by NIKA, NIKA2 is a completely new instrument, whose development has implied changes at all levels, from the detector arrays themselves all the way up to the optical chain of the 30m telescope.
\subsection{Optics}
To maximize the dimensions of the correct FoV, all the optical elements following the telescope's primary mirror (M1) and subreflector (M2) have been modified. The M3 and M4 mirrors, which are common to all the instruments of the cabin, have been replaced (the latter being used to select whether heterodyne or the continuum camera shall be used for observations). Two new dedicated mirrors (M5 and M6) have also been made, to send the light to the cryostat window.
Inside the cryostat, two additional cold ($\sim$80K) mirrors (M7 and M8) are present, followed by HDPE lenses on the coldest stages (1K and below). The band splitting is achieved by means of a dichroic installed at the 0.1K stage. The light in the 260GHz band is then further split by a wire-grid polarizer, that separates its horizontal and vertical components. This enables us to make polarization sensitive observations with a very low cross-pol level. When carrying out this kind of observations, a rotating half-wave plate is placed in front of the cryostat window to modulate the polarized light.

All optical elements have already been installed, and the cold optics have been extensively used for lab tests. The only exception is the rotating  half-wave plate, which is in the final fabrication steps and will be tested on-site during the first technical run.

\subsection{Cryostat}

\begin{figure}[b]
\begin{center}
\includegraphics[%
  width=0.75\linewidth, keepaspectratio]{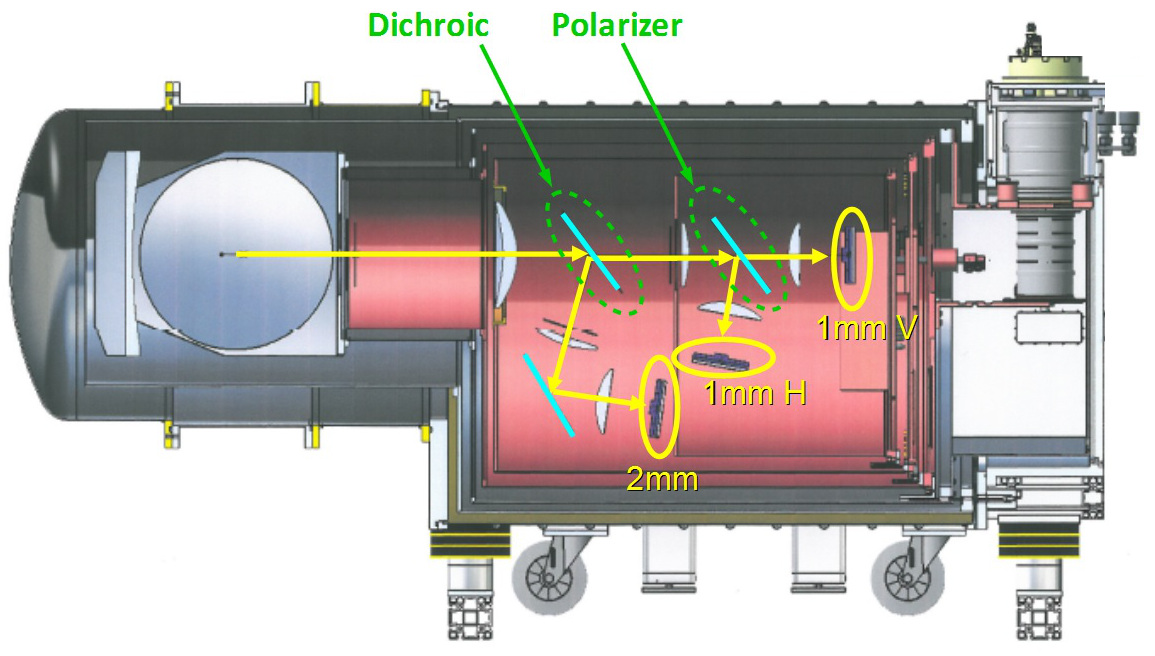} 
\end{center}
\caption{Schematic representation of the NIKA2 cryostat, with the optical path to the three arrays in evidence. The cryostat is 2.3m long and weighs more than 1 ton.}
\label{fig1}
\end{figure}

The NIKA2 cryostat is precooled to $\sim$5K by two Pulse Tubes (PT) working in parallel. A dilution refrigerator is then used to reach the base working temperature of about 150mK, which is achieved after 5 days. The system, which is completely cryogen free and can be fully remote controlled, can then be kept cold indefinitely.
The PT have remote motors, which are connected to the cryostat body using flexible tubes and rubber dampers. This suppresses the vibrations they induce, that can otherwise add a strong $1/f$ noise contribution to the detectors noise. To protect the detectors also from the effects of external magnetic fields, we added high permeability materials at each stage: one Mumetal cylinder at 300K, a Cryoperm one at 4K, a layer of Metglas tape at 100K and 50K, and a superconducting Aluminum screen at 200mK.

\subsection{Electronics}

The multiplexed readout of KID is achieved using dedicated electronics boards  called NIKELv1\cite{Bourrion}. These board can excite and readout up to 400 pixels over a 500MHz bandwidth, and apply a Direct Down Conversion method on the acquired output signal to determine the variations of amplitude and phase of each tone. 11 boards were available during the first technical run, out of the 20 needed for the final, complete configuration.

For the readout, the modulation technique developed for NIKA1 will be used\cite{Calvo}. This allows us to evaluate the variation of the resonant frequency $f_0$ with high precision, leading to very good photometric accuracy. The expected error on the absolute calibration of our detectors is below $10\%$, as already achieved with NIKA1.

\section{Detectors design and performance}

The NIKA2 pixels are based on Hilbert type LEKID\cite{Doyle08}$^,$\cite{Roesch12}. Thanks to the fractal shape of their inductor, such pixels can efficiently absorb both polarizations. The detectors are fabricated using thin (18$\textup{--}$25nm) Aluminum films, in order to increase the kinetic inductance fraction of the superconductor and make the KID more responsive. The thin Al film also increases the normal state surface resistivity, making it easier to match the free space impedance and directly absorb the radiation in the inductor.

\begin{figure}[t]
\begin{center}
\includegraphics[%
  width=0.65\linewidth, keepaspectratio]{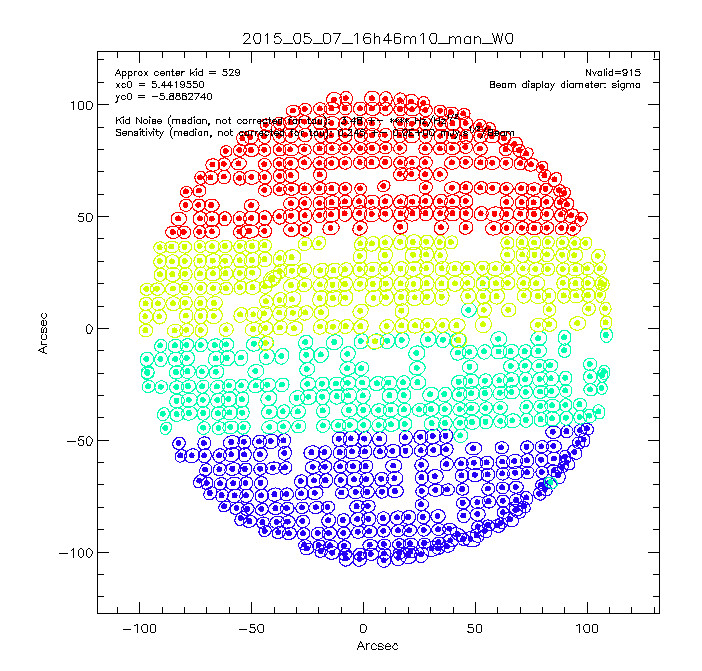} 
\end{center}
\caption{Reconstructed focal plane geometry of the 2mm CPW array. Each circle corresponds to one pixel, and the different colors are associated to the different readout feedlines. We excluded from this plot pixels having a too large noise or too elliptic beams. Even excluding such pixels, the yield obtained is above 80$\%$.}
\label{fig2}
\end{figure}

The focal plane has a diameter of $\sim 80$mm, and each array is fabricated on a single 4$\mathrm{''}$ HR Silicon wafer. More details on the fabrication process can be found elsewhere in these proceedings\cite{Goupy}$^,$\cite{Adane}. To maintain the telescope angular resolution, the focal plane sampling must be $\leq 1F\lambda$. We have tested different approaches, with a pitch between pixels varying from $0.7$ to $1F\lambda$, which corresponds to $2.3\textup{--}2.8$mm at 150GHz and $1.6\textup{--}2$mm at 260GHz, for a total pixels count of $600\textup{--}1000$ and $1200\textup{--}2000$ respectively. Although each NIKEL board can readout up to 400 pixels, for practical reasons the actual multiplexing factor is kept at a lower value of $150\textup{--}250$, so that 4 to 8 readout lines are needed per array.

We have investigated two different feedline geometries: Coplanar Waveguide (\emph{CPW}) and Microstrip (\emph{MS}). The CPW solution is more widespread and has already been tested in NIKA1. The main advantage of such an approach is that a back-illumination configuration can be adopted, using the substrate as an impedance matching Anti-Reflection (\emph{AR}) layer. Yet, the onset of spurious propagation modes can lead to the presence of standing waves along the feedline which, in turn, result in a large scatter of the coupling of the different resonators to it. This effect can be mitigated by implementing a series of bonding across the line. A simpler and more effective solution is that of recurring to a MS feedline, whose geometry is such that no spurious modes exist. The coupling of the detectors to the line and hence their performance are thus more uniform and predictable. On the other hand, the backside of the wafer is in this case metallized, in order to act as a groundplane, and the pixels must be forcefully front-illuminated. Using thin Al films and choosing the appropriate substrate thickness it is nonetheless possible to have $>40\%$ absorption efficiency over a $\sim20\%$ bandwidth, a value which is good enough for the aims of the NIKA2 experiment.

To characterize the performance of our arrays under conditions similar to those found at the telescope we use a \emph{sky simulator}. This consists of a copper flange, covered with a layer of carbon-loaded Stycast to make it almost completely black at millimetric wavelengths. This simulator is cooled using a single stage PT to $\sim30K$. Its optical emission is thus similar to that of the atmosphere which, under typical observing condition, corresponds to a grey body with $\sim10\%$ emissivity and a temperature of $\sim270K$. A metallic sphere of 4mm diameter can be moved in front of the sky simulator, thus mimicking the passage of a planet comparable to Uranus. This allows us to reconstruct an image of the focal plane and evaluate the responsivity of each pixel.

The baseline arrays for the first technical run (October 2015) have already been chosen. For the 150GHz band, a 1040 pixels array with CPW feedline and a sampling of $0.7F\lambda$ will be used. The back of the wafer has been diced to improve its effectiveness as an AR layer. The 260GHz band will be based on two MS arrays, each containing 1200 pixels, sampling the focal plane at $1F\lambda$. The three arrays have been fully characterized using the sky simulator. The results are summarized in table \ref{tableRes}.

\begin{figure}[t]
\begin{center}
\includegraphics[%
  width=0.85\linewidth, keepaspectratio]{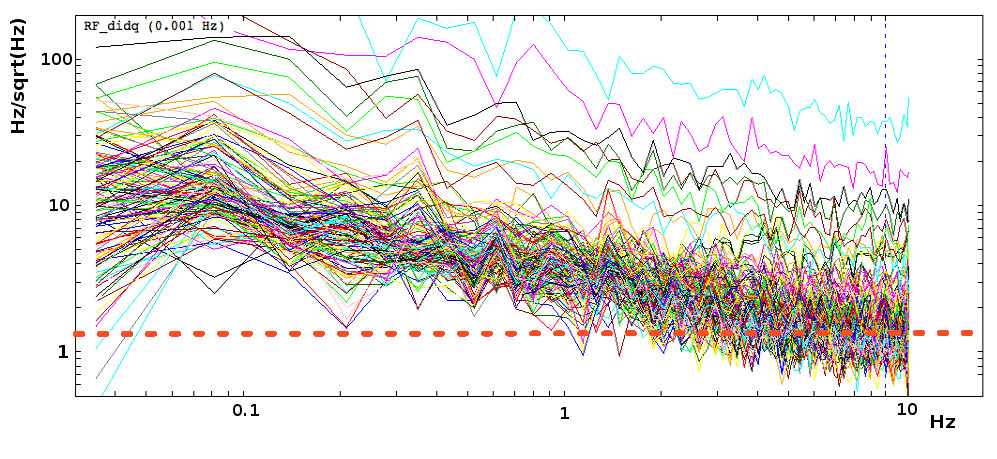} 
\end{center}
\caption{Noise spectra of the pixels of one the baseline 150GHz array feedlines, obtained under an optical load representative of the one expected at the IRAM 30m site. Using the same pixels, we measured a shift of the resonant frequency of about 3kHz for a change of 3.6K in the sky simulator temperature. Combining this two data it is possible to evaluate the Signal-to-Noise Ratio (SNR) of each pixel. The noise at low $f$ is mostly correlated across all the pixels, and can thus be efficiently removed using an appropriate decorrelation technique during data analysis.}
\label{fig3}
\end{figure}

The per pixel performance obtained in the lab are in line with those found for the NIKA detectors when carrying the same kind of tests. For NIKA, we demonstrated an on-sky sensitiviy on point-like sources of 10 and 25mJy$\cdot$s$^{1/2}$ at 150 and 260GHz respectively in good observing conditions. We therefore expect to obtain similar values for the NIKA2 arrays. Thus, considering the larger FoV, NIKA2 will grant a factor $\sim10$ increase in mapping speed with respect to its predecessor, making it one of the most effective cameras worldwide at these wavelengths.

\begin{figure}[!b]
\begin{center}
\includegraphics[%
  width=0.85\linewidth, keepaspectratio]{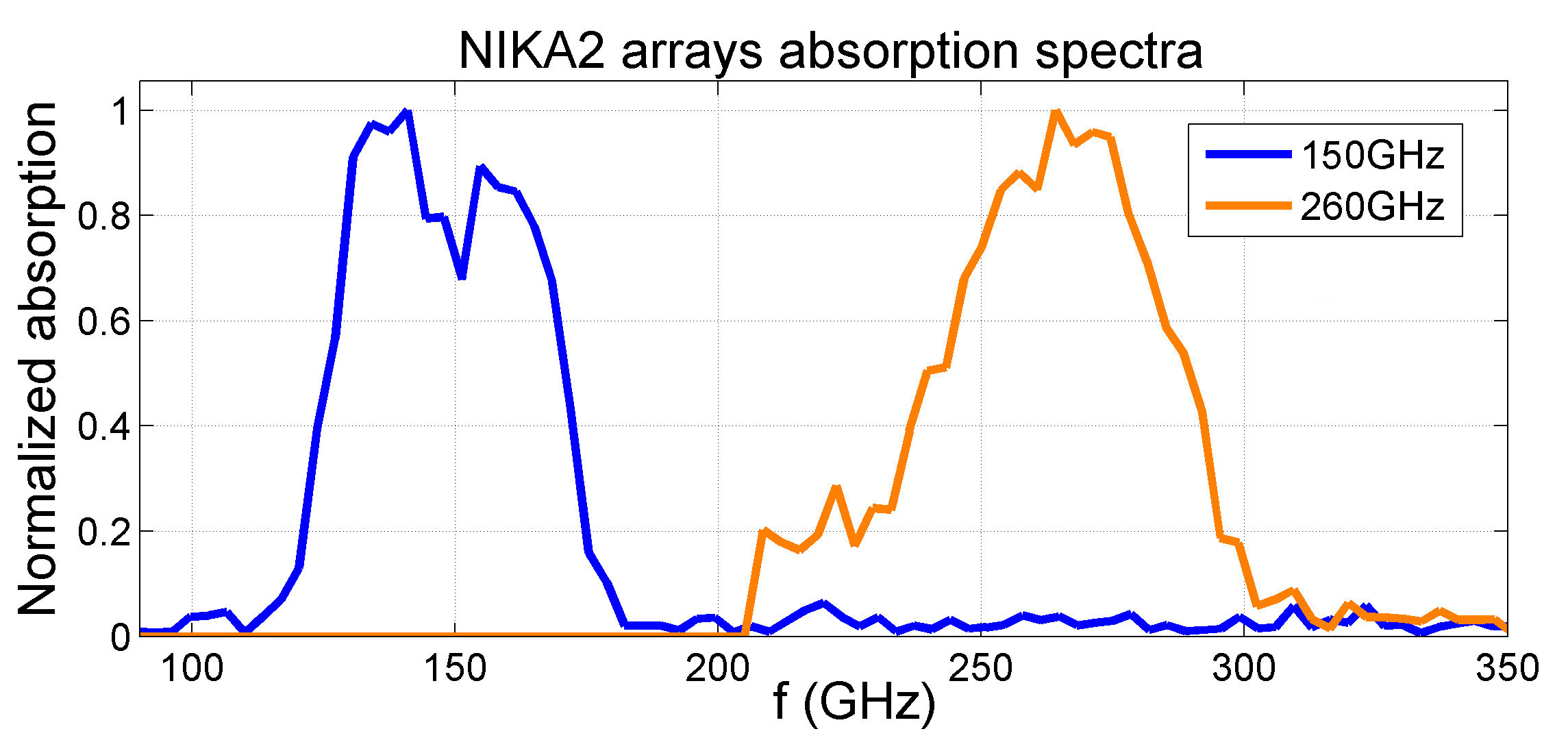} 
\end{center}
\caption{Absorption spectra of two sample pixels of the NIKA2 arrays: in blue the pixel of the 150GHz array, and in orange that of one of the 260GHz arrays. The absorption in each band has been normalized to 1. The 150GHz CPW array has a flatter absorption band thanks to the use of the substrate as an AR layer.}
\label{fig4}
\end{figure}
The effective bandpass of the different channels is determined by the product of the transmissivity of the optical filters and of the absorption efficiency of the corresponding pixels. We measure it using a Martin-Pupplet Interferometer (MPI), with which we can determine the shape of the absorption spectrum with a resolution of about 3GHz. Although the absolute value of the total absorption is difficult to know precisely, as it depends upon the details of the optics and of the MPI components, its shape is all that is needed to perform accurate photometric measurements. The absorption spectra of two of the baseline arrays are shown in Fig. \ref{fig4}. The choice of the substrate thickness and the use of $\lambda/4$ backshorts allows us to fine tune the position of the absoprtion band, which is chosen in order to take advantage of the atmospheric windows available, while at the same time avoiding the contamination due to the emission lines of molecules, in particular water vapour and ozone.

Until the end of 2016, NIKA2 will be devoted to technical runs for its commissioning. During this time, new arrays will be fabricated and tested. If major performance improvements were to be obtained, access will be granted to the camera for the replacement of the current baseline arrays.



\begin{table}[t]
\centering
\begin{tabular}{|c|c|c|c|}
\hline
Array & Noise @ 1Hz & Responsivity & SNR @ 1Hz\\
 &  (Hz/Hz$^{0.5}$)  & (kHz/K) &(mK/Hz$^{0.5}$)\\
\hline
150GHz & 1$\textup{--}$2 & 0.8 & $\sim$1.5 \\
260GHz H & 2$\textup{--}$3 & 2 & $\sim$1.2  \\
260GHz V & 2$\textup{--}$3 & 2 & $\sim$1.2 \\
 \hline 
\end{tabular}
\caption{Overview of the performance of the baseline arrays of NIKA2. The reported values account for about $80\%$ of the total pixels, the remaining $20\%$ being either dead pixels or pixels that have abnormally high noise levels.}
\label{tableRes}

\end{table}

\section{Conclusions}

\if
NIKA2, the new photometric instrument of the IRAM 30m telescope, is now ready for installation. The development of the instrument took three years from its first conceptual designs to the final assembly. This confirms how LEKID detectors couple good performance to a relatively easy implementation, and are today a very good solution for millimeter wave detection.

The main goal of the NIKA2 project was that of fully sampling the 6.5arcmin correct FoV available at the telescope. To achieve this, we have built a completely new cryostat, and replaced all the optical elements  

\fi

NIKA2 is the next generation (2015-2025) resident photometric instrument of the IRAM 30m telescope. Over the last three years, we have fabricated all the needed components. This include, in particular, a dry cryostat with a dilution refrigerator stage, that has already been cooled down multiple times to a base temperature of 150mK. To maximize the available FoV, we have made a completely new optical design and have replaced all the components of the optical chain that follow the secondary mirror. The correct FoV is now 6.5arcmin in diameter.

In order to fully sample such a large FoV, we have increased the pixel count  by a factor $\sim10$ with respect to NIKA, all while keeping similar per pixel performance, as they already approached the photon noise limit of the Sierra Nevada site. NIKA2 will thus lead to a tenfold increase of the mapping speed, and will also allow for the measurement of the linear polarization of light in the 260GHz band.

With all the subsystems built and validated, NIKA2 has been installed at the telescope during a dedicated technical run that took place in October 2015. This will be followed by a one-year commissioning period, after which NIKA2 will be made available to the external astronomers. The unique features of this camera will then open the way to the exploration of new and exciting scientific cases.



\if

From the detectors point of view, the larger FoV with respect to NIKA, we have therefore increased the pixel count by a factor $\sim10$, all while keeping similar per pixel performance, which already approached the photon noise limit of the Sierra Nevada site. NIKA2 will thus lead to a ten folds increase of the mapping speed, and will also allow for the measurement of the linear polarization of light in the 260GHz band.

With all the subsystems built and validated, NIKA2 is now ready for final installation and assembly at the telescope. A dedicated technical run will take place in October 2015, and will be followed by a one-year commissioning period. After this, NIKA2 will be made available to the external astronomers. The unique features of this camera will then open the way to the exploration of new and exciting scientific cases.


The NIKA dual-band continuum instrument is now fully operational and will be
open to the 30m telescope observers from the next Winter (2013/14). In six technical and scientific (restricted to the NIKA collaboration) runs we demonstrated
competitive sensitivities at both 150 GHz and 260 GHz and good photometry
performance. Moreover, thanks to the intrinsic KID linearity we provide the astronomers with a real-time estimation of the atmospheric opacity correction. This
correction is calculated along the line-of-sight and for the real detectors band. It
is thus in the long term more reliable than the IRAM tipping radiometer which
works at 225 GHz at a fixed azimuth direction.
The next generation NIKA-2 camera will cover a larger field-of-view (6.5 arcmin compared to 2 arc-min in NIKA), preserve dual-band imaging capabilities
and measure in addition the linear polarisation at 260 GHz. This is achieved with
three large LEKID arrays (sensitive area diameter = 80 mm) and a total pixels
count of 5,000. NIKA-2, being fabricated in Grenoble, is officially selected by
IRAM as the next generation continuum instrument at the 30m telescope. It will
be installed for Commissioning in 2015.
\fi

\begin{acknowledgements}
This work has been partially funded by the ANR under the contract ”NIKA2”, and has been partially supported
by the LabEx FOCUS ANR-11-LABX-0013. 
We would like to thank the IRAM staff for their excellent support during the campaign. 
This work has been partially funded by the Foundation Nanoscience Grenoble, the ANR under the contracts "MKIDS" and "NIKA". 
This work has been partially supported by the LabEx FOCUS ANR-11-LABX-0013. 
This work has benefited from the support of the European Research Council Advanced Grant ORISTARS under the European Union's Seventh Framework Programme (Grant Agreement no. 291294).
The NIKA2 cryostat and the readout electronics have
been designed and assembled by the Cryogenics and Electronics groups in Grenoble. In particular, we acknowledge the key contributions of P.E. Wolf, A. Gerardini, G. Donnier-Valentin, H. Rodenas, O. Exshaw, G. Simiand, C. Vescovi, E. Perbet and S. Roudier.
R. A. would like to thank the ENIGMASS French LabEx for funding this work. 
\end{acknowledgements}

\end{document}